\documentclass[aps,prl,preprint,groupedaddress]{revtex4}

\usepackage{graphicx}

\begin{document}

\title{Aharonov-Bohm signature for neutral excitons in type-II
quantum dot ensembles}

\author{E. Ribeiro}
\email{evaldo@lnls.br}
\author{G. Medeiros-Ribeiro}
\author{W. Carvalho Jr.}
\affiliation{Laborat\'orio Nacional de Luz S\'{\i}ncrotron, PO Box
6192, 13084-971 Campinas - SP, Brazil}

\author{A. O. Govorov}
\affiliation{Ohio University, Department of Physics and
Astronomy, Clippinger Research Labs, Athens OH 45701, USA}

\date{\today}

\begin{abstract}

The Aharonov-Bohm (AB) effect is commonly believed to be a typical
feature of the motion of a charged particle interacting with the
electromagnetic vector potential. Here we present a
magnetophotoluminescence study of type-II InP/GaAs self-assembled
quantum dots, revealing the Aharonov-Bohm-type oscillations for
neutral excitons when the hole ground state changes its angular
momentum from $\ell_{h}$ = 0 to $\ell_{h}$ = 1, 2, and 3. The hole
ring parameters derived from a simple model are in excellent
agreement with the structural parameters for this system.
\end{abstract}

\pacs{71.35.Ji, 78.67.Hc, 73.22.Dj}

\maketitle

In 1959, Aharonov and Bohm proposed an experiment to assess the
manifestations of the electromagnetic potentials in the quantum
domain \cite{aharonov59}. By allowing a charged particle to
circulate a confined magnetic field flux region, they showed that
surprisingly, there exist effects of the vector potential on the
charged particles moving outside. After the circulation the
particle wavefunction acquires a phase that is proportional to the
magnetic flux contained within the closed path \cite{berry84}.
Therefore, all observable phenomena depend upon the flux $\Phi$
through the excluded region, and are shown to be periodic with
period $\Phi_0 = hc/e$ \cite{byers61,bloch68,buttiker83}. This
oscillatory characteristic is the signature of the Aharonov-Bohm
(AB) effect \cite{aharonov59,peshkin89}.

The definite measurement of the original AB effect was performed
by Tonomura {\it et al.} in a very clever experiment
\cite{tonomura86,peshkin89}. Investigation and applications of the
AB concepts in condensed matter have been carried out in metals
\cite{webb85} and superconductors \cite{byers61,bloch68,parks64}.
With the advance of lithography and semiconductor growth
techniques, it was possible to devise mesoscopic
\cite{timp87,ford88} and nanoscopic \cite{fuhrer01,yau02} devices
where many theoretically predicted signatures evidencing ring-like
electronic properties \cite{altshuler81,buttiker85,aronov87} have
been observed by transport experiments. For the case of few
electrons, self-assembled quantum rings \cite{lorke00} were used
for the observation of the characteristic spectra including
excitations of states of different angular momentum
\cite{lorke00}. Due to the small sizes of these structures, it was
possible for the carriers to maintain the coherence of their
wavefunctions during their motion (at sufficiently low
temperatures), thus allowing the observation of relative quantum
phase effects (AB-like).

It was proposed that a neutral excitation could also exhibit such
an AB oscillatory behavior \cite{chaplik95,kalameitsev98}.
Electron-hole pairs created by optical excitation and bound
together via Coulomb interaction (excitons) are a good candidate
when either one or both carriers are confined to a ring-like
geometry. The existence of the AB oscillations for neutral
excitons in semiconductor quantum rings, where both carriers are
confined in the rings, has been a matter of controversy in recent
years \cite{chaplik95,romer00,song01,hu01,govorov02a,govorov02b}
since some of the work predicted that the AB oscillations might be
very weak because of exponentially small electron-to-hole tunnel
amplitude \cite{chaplik95,kalameitsev98} or would vanish for a
finite width ring \cite{song01,hu01}. Recently, the AB effect was
observed in optical experiments by Bayer {\it et al.} on
quantum-rings fabricated by lithographical methods \cite{bayer03}.
They recorded AB oscillations for {\it charged} excitons. Perhaps
due to the aforementioned reasons, the AB effect was not found in
the spectra of {\it neutral} excitons. The second possibility for
observing AB-like oscillations in the optical spectra of a neutral
excitation is the case of type-II quantum dots (QD)
\cite{kalameitsev98, govorov02a,janssens01,janssens02}, where the
confinement of one carrier inside the QD and the other carrier in
the barrier naturally creates a ring-like structure (see Fig. 1).
In such a structure, the carrier outside the QD would be bound to
that confined in the QD via Coulomb attraction. This spatial
charge separation produces a polarization of the exciton and thus
AB oscillations could be observed in the energy of those carriers
that are confined in the ring-shaped orbit around the QD.
Moreover, the AB effect with a {\it polarized} exciton may not
involve an electron-to-hole tunnel amplitude
\cite{kalameitsev98,govorov02a}. There are several self-assembled
quantum dot systems where a type-II heterojunction is expected;
for instance, GaSb on GaAs \cite{hatami95} and Ge on Si
\cite{schittenhelm95} provide a three-dimensional confinement for
the holes, thereby creating a ring-like potential for the
electrons. Alternatively, InP on InGaP \cite{hayne00} and InP on
GaAs \cite{ribeiro02} provide the localization of the electrons in
the QD, and consequently a ring-like hole wavefunction. Figure 1
sketches the valence band potential in a very simplified way, not
taking into account the contribution of the strain. It is known
that the QD strains the GaAs substrate and cap layer, and that in
turn changes the potential landscape in the GaAs, creating an
additional ring-like potential around the QD for heavy-hole
confinement \cite{xu91}. The light hole may have a well-like
strain-induced potential below the QD \cite{xu91}, however being
shallower than the heavy hole ring potential. InP QDs are
particularly suitable for observing oscillatory behavior with AB
period $\Phi_{0}$. With diameters of the order of 35 nm one
obtains a period of few Tesla for the AB effect.

\begin{figure}
\includegraphics{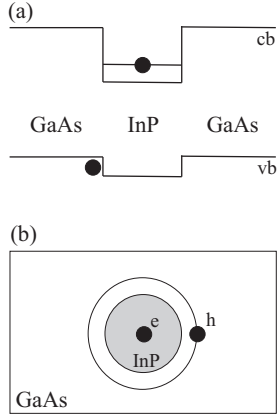}%
\caption{\label{f1} Sketches of the type-II InP/GaAs QDs: (a)
conduction and valence band profiles, indicating the spatial
separation of electrons and holes; (b) top view of the QD plane,
indicating the holes confined to a ring around the QD due to the
Coulomb interaction with the electron trapped into the dot.}
\end{figure}

Here we present data on InP/GaAs self-assembled quantum dot
ensembles where the confined magnetoexcitons trap fluxes in the
range of 0-3 flux quanta, corresponding to a magnetic field span
of 0-12 T. Samples were grown at different rates and PH$_{3}$
fluxes by metal-organic chemical vapor deposition in a commercial
reactor on nominally flat semi-insulating Cr-doped GaAs [001]
substrates at 550 $^{o}$C \cite{ribeiro02}. The structures
consisted of a 300 nm undoped GaAs grown at 600 $^{o}$C buffer
layer followed by the QD layer. The islands were then capped with
a 50 nm undoped GaAs cap. For this work, the sample with the
narrowest luminescence line width ($\sim$35 meV) was chosen,
representing an ensemble of $3 \times 10^{10}$ QDs/cm$^{2}$
\cite{ribeiro02} with small size dispersion. The dimensions of the
QDs were determined by cross-section transmission electron
microscopy (TEM) \cite{maltez03}, and were found to be (32 $\pm$
6) nm and (4 $\pm$ 2) nm in average diameter and height. The InP
QDs are not cylindrically symmetric, being elongated in the [110]
direction. The photoluminescence (PL) experiments were performed
at 2 K, with magnetic fields up to 12 T (in 0.2 T steps), and
using an Ar laser with 2 W/cm$^{2}$ as excitation source. These
conditions guaranteed the filling of only the system ground state,
allowing the assessment of single particle energies. For higher
excitation power a line broadening and luminescence shift
\cite{nakaema02} takes place, indicating excited state population,
which would complicate our analysis.

Fig. 2 shows the InP QDs PL spectra for $B = 0$ (circles) and $B =
12$ T (triangles). The total energy shift within this field range
was 4.5 meV. The PL peak intensity increased about 50\% from 0 to
12 T, indicating an enhancement of the electron and hole
wavefunction overlap. The spectra for all the field values were
fit with a single Gaussian line shape in order to extract the
evolution of the PL peak position of the InP QDs as a function of
the applied magnetic field, $E_{PL}(B)$, which is plotted in Fig.
3(a). There one can see that instead of the typical monotonic
diamagnetic energy shift characteristics of type-I QD
\cite{wang96}, the InP QD PL shows an oscillatory behavior in the
$E_{PL}(B)$ curve, very similar to what has been predicted by
Kalame\v{\i}tsev and collaborators \cite{kalameitsev98}. In that
work, it was expected that the maxima of oscillating part in
$E_{PL}(B)$ would develop every time that the lower-lying carrier
state changed its angular momentum from $\ell_{e}$ = 0 to
$\ell_{e}$ = $-1$ and so forth. These transitions were seen to
occur when $R^{2}/l_{B}^{2} \sim |\ell_{e}|$ \cite{kalameitsev98},
with $R$ being the ring radius, $l_{B}^{2} = \hbar/eB$ the
magnetic length and $\ell_{e}$ the electron state angular
momentum. Rewriting this expression for the holes in InP QDs, $R
\sim (\hbar \ell_{h}/eB)^{1/2}$ and then by inserting the
experimental values of $B$ for the observed plateau-like regions
[Figure 3(a)] an estimate for the hole ring radius can be
obtained. Averaging the results for the three plateaux, a radius
of $R = (16 \pm 1)$ nm is obtained, in agreement to what has been
measured by TEM \cite{maltez03}.

\begin{figure}
\includegraphics{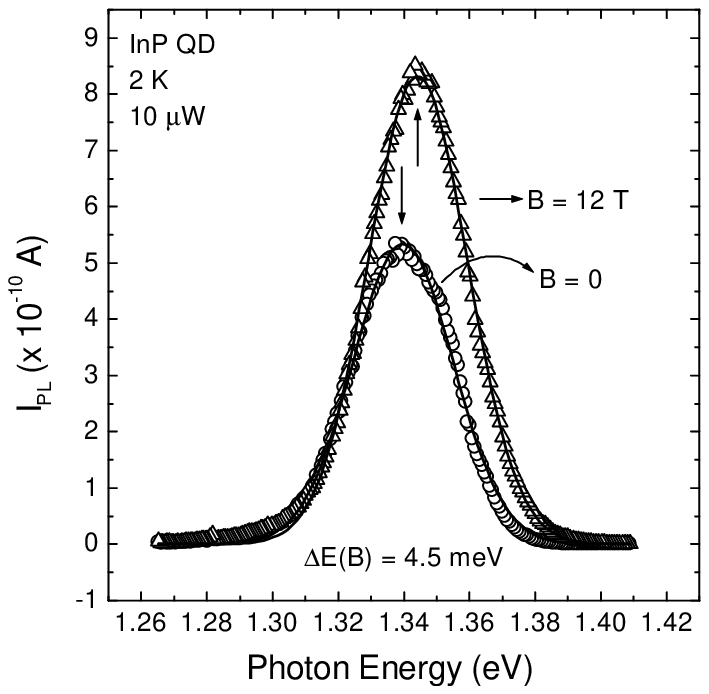}%
\caption{\label{f2} PL spectra of the InP QDs for $B = 0$
(circles) and $B = 12$ T (triangles). Solid lines are Gaussian
fits to the data. The total energy shift is 4.5 meV for 12 T.}
\end{figure}

Nevertheless, it is possible to further elaborate the analysis of
the data. The electron ground state of InP/GaAs QDs has been
investigated by capacitance spectroscopy (CV) \cite{bufon04}. It
has already been established \cite{wang96} that the lateral
confinement for the self-assembled quantum dot can be reasonably
modelled by a parabolic potential. The diamagnetic shift of the
electron ground state can be described by $E_{e} = ((\hbar
\omega_{0})^{2} + (\hbar \omega_{c})^{2}/4)^{1/2}$ \cite{fock28},
with $\omega_{c}$ being the cyclotron frequency. For InP/GaAs QDs
grown under identical nominal conditions, a characteristic energy
of $\hbar \omega_{0} = (5.6 \pm 0.9)$ meV was inferred
\cite{bufon04}. Bearing this information in mind, an expression
for $E_{PL}(B)$ can be written, consisting of three terms: a
constant contribution, related to the energy gap of the material
and the exciton binding energy; the diamagnetic energy shift for
the electrons, discussed above; and the variation of the hole
energy with the magnetic field \cite{govorov02b,janssens01}:

\begin{equation}
E_{PL}(R,B) = E_{g} + \sqrt{(\hbar \omega_{0})^{2} +
\left(\frac{\hbar \omega_{c}}{2}\right)^{2}} +
\frac{\hbar^{2}}{2m^{*}_{h}R^{2}} \left(\ell_{h} -
\frac{\Phi}{\Phi_{0}}\right)^{2},
\end{equation}

where $m^{*}_{h}$ is the hole effective mass, $\Phi_{0} = hc/e$ is
the flux quantum and $\Phi = \pi R^{2} B$ is the total magnetic
flux through the ring of radius $R$. Using the above result for
$\hbar \omega_0$, both the constant and the electron contributions
in the experimental data [see solid curve in Fig. 3(a)] can be
subtracted. The result is shown in Fig. 3(b), where:

\begin{equation}
E_{h}(R,B) = E_{PL}(R,B) - E_{g} - E_{e} =
\frac{\hbar^{2}}{2m^{*}_{h}R^{2}} \left(\ell_{h} -
\frac{\Phi}{\Phi_{0}}\right)^{2}.
\end{equation}

As shown in Fig 3(b), $E_{h}$ (open squares) clearly oscillates as
a function of the magnetic field, indicating phase coherence for
the hole wavefunctions. Since $\Phi_{0}$ is a constant (the
Aharonov-Bohm period of the excitonic energy oscillations
\cite{aharonov59,buttiker85,peshkin89}), the experimental data
were fit with the parabolas given by Eq. (2), one for each value
of $\ell_{h}$ = 0, 1, 2, and 3, indicating the number of flux
quanta enclosed by the hole ring. The result is shown by the
dashed curves in the upper part of Fig. 3(b), from which $R =
(19.1 \pm 0.4)$ nm and $\Phi_{0}$ = 3.61 T were obtained. One
should keep in mind that the hole ring must have a radius slightly
larger than the QD diameter (Fig. 1), and therefore the obtained
values are in excellent agreement with the structural and
electronic results discussed above \cite{maltez03}.

\begin{figure}
\includegraphics{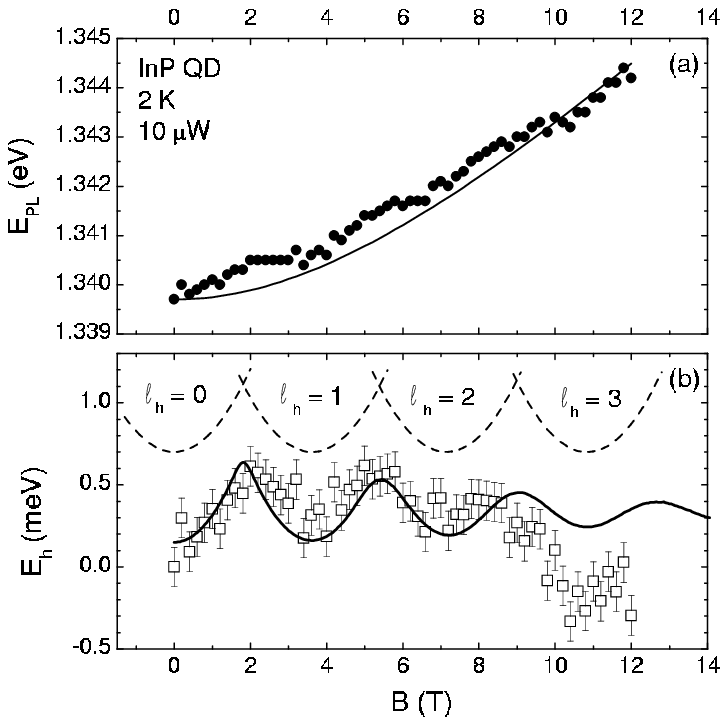}%
\caption{\label{f3} (a) The PL peak position as a function of the
applied magnetic field (solid circles). Three plateau-like
structures are clearly observed. The error bars are smaller than
the solid circles. (b) Hole energy dependence on the magnetic
field (open squares), showing the Aharonov-Bohm oscillations with
period $\phi_{0}$. The error bars are the uncertainties on the
peak position after the Gaussian fits. The dashed curves are
parabolas following Eq. (2) shifted upward by arbitrary energy.
Calculated $E_{h}$ [Eq. (2)] for a QD ensemble described by a
Gaussian size distribution for $R$, with $R_{peak}$ = 19.1 nm and
full width at half maximum of $2 \Delta R_{CV}$ = 1.6 nm is shown
by the solid curve. Note the progressive offsets of the parabola
minima for increasing $\ell_{h}$.}
\end{figure}

It is worth noting that an ensemble of islands and not a single QD
was probed. Nevertheless, the size dispersion did not suppress the
AB oscillations. Some considerations can be done on the influence
of the size dispersion $\Delta R$ on the experimental results.
From CV experiments \cite{bufon04} an effective size distribution
of $\Delta R_{CV}$ = 0.8 nm was inferred for the
electron-confining potential. This CV estimate for $\Delta R$
would be a lower bound for the experimentally-probed size
dispersion. In order to model in a more quantitative fashion the
effect of the QD ensemble on the hole-ring energy oscillations,
$E_{h}(R,B)$ [eq. (2)] was calculated and weighted by a Gaussian
size distribution function, where $R$ determined in Fig. 3(b) is
the central peak radius and $2 \Delta R_{CV}$ is its full width at
half maximum. The result is plotted in Fig. 3(b) (solid curve). It
can be seen that the main effect of the size dispersion on $E_{h}$
is a progressive attenuation of the oscillation intensity, as one
would expect, and it is consistent with the decrease in the
amplitude of the data in Fig. 3(b). By comparing the data and the
solid curve in Fig. 3(b) one can see that the experimental data of
the hole ring are consistent with an effective size dispersion
obtained from $\Delta R_{CV}$.

Although describing the oscillation period and damping in a good
agreement with independent structural/electronic results, this
simplified model fails to predict the energy offset of the last
parabola (corresponding to $\ell = 3$). The contributions for the
offset might be: (i) the experimental uncertainty in $\hbar
\omega_{0}$ obtained from the CV fit [solid curve in Fig 3(a)];
(ii) a possible dependence of the exciton binding energy on the
magnetic field; (iii) deviations from the parabolic lateral
confinement due to the actual QD morphology. Clearly, a more
detailed and complete modelling of the system, including the
points indicated above, is needed for understanding all the
experimental features presented in Fig. 3(b), but such a task is
beyond the scope of the present work. Nevertheless, the simple
model used here consistently describes the main physical issues
connected to the observation of the hole AB oscillations (Fig. 3).

In summary, we have presented clear evidence of Aharonov-Bohm
oscillations in the spectrum of electrically neutral excitons
confined in type-II InP/GaAs quantum dots. We have observed the
trapping of up to three flux quanta inside the ring trajectory of
the hole, for a reasonable span of magnetic fields. Despite the
presence of a size dispersion, the AB oscillations were easily
observable, evidencing the phase coherence for holes moving around
QDs.


\begin{acknowledgments}
The magnetic field experiments were performed at the Optical
Properties Group facilities at Campinas State University
(Unicamp). We would like to thank H. Gazetta Filho for the
technical support on sample growth. We gratefully acknowledge
financial support from Funda\c{c}\~{a}o de Amparo \`{a} Pesquisa
do Estado de S\~{a}o Paulo and MCT-CNPq.
\end{acknowledgments}


\begin{thebibliography}{99}

\bibitem{aharonov59} Y. Aharonov and D. Bohm, {\it Phys. Rev.}
{\bf 115}, 485 (1959).

\bibitem{berry84} M. V. Berry, {\it Proc. R. Soc. Lond. A}
{\bf 392}, 45 (1984).

\bibitem{byers61} N. Byers and C. N. Yang, {\it Phys. Rev. Lett.}
{\bf 7}, 46 (1961).

\bibitem{bloch68} F. Bloch, {\it Phys. Rev. Lett.} {\bf 21}, 1241
(1968).

\bibitem{buttiker83} M. B\"{u}ttiker, Y. Imry, and R. Landauer,
{\it Phys. Lett.} {\bf 96A}, 365 (1983).

\bibitem{peshkin89} For a review on the subject, see M. Peshkin and
A. Tonomura, ``The Aharonov-Bohm effect'', in {\it Lecture Notes
in Physics} {\bf 340} (Springer-Verlag, Berlin 1989).

\bibitem{tonomura86} A. Tonomura, N. Osakabe, T. Matsuda, T.
Kawasaki, J. Endo, S. Yano, and H. Yamada, {\it Phys. Rev. Lett.}
{\bf 56}, 792 (1986).

\bibitem{webb85} R. A. Webb, S. Washburn, C. P. Umbach, and R. B.
Laibowitz, {\it Phys. Rev. Lett.} {\bf 54}, 2696 (1985); S. Washburn
and R. A. Webb, {\it Rep. Prog. Phys.} {\bf 55}, 1311 (1992).

\bibitem{parks64} R. D. Parks and W. A. Little, {\it Phys. Rev.}
{\bf 133}, A97 (1964).

\bibitem{timp87} G. Timp, A. M. Chang, J. E. Cunningham, T. Y.
Chang, P. Mankiewich, R. Behringer, and R. E. Howard, {\it Phys.
Rev. Lett.} {\bf 58}, 2814 (1987).

\bibitem{ford88} C. J. B. Ford, T. J. Thornton, R. Newbury, M.
Pepper, H. Ahmed, C. T. Foxon, J. J. Harris, and C. Roberts, {\it
J. Phys. C: Solid State Phys.}, {\bf 21}, L325 (1988).

\bibitem{fuhrer01} A. Fuhrer, S. L\"{u}scher, T. Ihn, T. Heinzel, K.
Ensslin, W. Wegscheider, and M. Bichler, {\it Nature} {\bf 413},
822 (2001); ibid., {\it Microelectr. Eng.} {\bf 63}, 47 (2002).

\bibitem{yau02} J.-B. Yau, E. P. De Poortere, and M. Shayegan,
{\it Phys. Rev. Lett.} {\bf 88}, 146801 (2002).

\bibitem{altshuler81} B. L. Al'tshuler, A. G. Aronov, and B. Z.
Spivak, {\it Pis'ma Zh. \'Eksp. Teor. Fiz.} {\bf 33}, 101 (1981)
[{\it JETP Lett.} {\bf 33}, 94 (1981)].

\bibitem{buttiker85} M. B\"{u}ttiker, Y. Imry, R. Landauer, and
S. Pinhas, {\it Phys. Rev. B} {\bf 31}, 6207 (1985).

\bibitem{aronov87} A. G. Aronov and Yu. V. Sharvin, {\it Rev. Mod.
Phys.} {\bf 59}, 755 (1987).

\bibitem{lorke00} A. Lorke, R. J. Luyken, A. O. Govorov, J. P.
Kotthaus, J. M. Garcia, and P. M. Petroff, {\it Phys. Rev. Lett.}
{\bf 84}, 2223 (2000).

\bibitem{chaplik95} A. V. Chaplik, {\it Pis'ma Zh. \'Eksp. Teor.
Fiz.} {\bf 62}, 885 (1995) [{\it JETP Lett.} {\bf 62}, 900 (1995)].

\bibitem{kalameitsev98} A. B. Kalame\v{\i}tsev, V. M. Kovalev,
and A. O. Govorov, {\it Pis'ma Zh. \'Eksp. Teor. Fiz.} {\bf 68},
634 (1998) [{\it JETP Lett.} {\bf 68}, 669 (1998)].

\bibitem{romer00} R. A. R\"{o}mer and M. E. Raikh, {\it Phys.
Rev. B} {\bf 62}, 7045 (2000).

\bibitem{song01} J. Song and S. E. Ulloa, {\it Phys. Rev. B}
{\bf 63}, 125302 (2001).

\bibitem{hu01} H. Hu, J.-L. Zhu, D.-J. Li, and J.-J Xiong,
{\it Phys. Rev. B} {\bf 63}, 195307 (2001).

\bibitem{govorov02a} A. O. Govorov, A. V. Kalame\v{\i}tsev, R.
Warburton, K. Karrai, and S. E. Ulloa, {\it Physica E} {\bf 13},
297 (2002).

\bibitem{govorov02b} A. O. Govorov, S. E. Ulloa, K. Karrai, and R.
J. Warburton, {\it Phys. Rev. B} {\bf 66}, 081309 (2002); S. E.
Ulloa, A. O. Govorov, A. V. Kalame\v{\i}tsev, R. Warburton, and K.
Karrai, {\it Physica E} {\bf 12}, 790 (2002); A. V.Maslov and D.
S. Citrin, {\it Phys. Rev. B} {\bf 67}, 121304 (2003).

\bibitem{bayer03} M. Bayer, M. Korkusinski, P. Hawrylak, T.
Gutbrod, M. Michel, and A. Forchel, {\it Phys. Rev. Lett.} {\bf
90}, 186801 (2003).

\bibitem{janssens01} K. L. Janssens, B. Partoens, and F. M.
Peeters, {\it Phys. Rev. B} {\bf 64}, 155324 (2001).

\bibitem{janssens02} K. L. Janssens, B. Partoens, and F. M.
Peeters, {\it Phys. Rev. B} {\bf 66}, 075314 (2002).

\bibitem{hatami95} F. Hatami, N. N. Ledentsov, M. Grundmann, J.
Böhrer, F. Heinrichsdorff, M. Beer, D. Bimberg, S. S. Ruvimov, P.
Werner, U. G\"osele, J. Heydenreich, U. Richter, S. V. Ivanov, B.
Ya. Meltser, P. S. Kop'ev, and Zh. I. Alferov, {\it Appl. Phys.
Lett.} {\bf 67}, 656 (1995).

\bibitem{schittenhelm95} P. Schittenhelm, M. Gail, J. Brunner, J.
F. N\"utzel, and G. Abstreiter, {\it Appl. Phys. Lett.} {\bf 67},
1292 (1995); A. I. Yakimov, N. P. Stepina, A. V. Dvurechenskii,
A. I. Nikiforov, and A. V. Nenashev, {\it Phys. Rev. B} {\bf 63},
045312 (2001).

\bibitem{hayne00} M. Hayne, R. Provoost, M. K. Zundel, Y. M.
Manz, K. Eberl, and V. V. Moshchalkov, {\it Phys. Rev. B} {\bf
62}, 10324 (2000); M. Sugisaki, H.-W. Hen, S. V. Nair, K. Nishi,
and Y. Masumoto, {\it Phys. Rev. B} {\bf 66}, 235309 (2002).

\bibitem{ribeiro02} E. Ribeiro, R. L. Maltez, W. Carvalho Jr., D.
Ugarte, and G. Medeiros-Ribeiro, {\it Appl. Phys. Lett.} {\bf
81}, 2953 (2002).

\bibitem{xu91} Z. Xu and P. M. Petroff, {\it J. Appl. Phys.}
{\bf 69}, 6564
(1991).

\bibitem{maltez03} R. L. Maltez, E. Ribeiro, W. Carvalho Jr., D.
Ugarte, and G. Medeiros-Ribeiro, {\it J. Appl. Phys.} {\bf 94},
3051 (2003).

\bibitem{nakaema02} M. K. K. Nakaema, F. Iikawa, M. J. S. P.
Brasil, E. Ribeiro, G. Medeiros-Ribeiro, W. Carvalho Jr., M. Z.
Maialle, and M. H. Degani, {\it Appl. Phys. Lett.} {\bf 81}, 2743
(2002).

\bibitem{wang96} P. D. Wang, J. L. Merz, S. Fafard, R. Leon, D.
Leonard, G. Medeiros-Ribeiro, M. Oestreich, P. M. Petroff, K.
Uchida, N. Miura, H. Akiyama, and H. Sakaki, {\it Phys. Rev. B}
{\bf 53}, 16458 (1996); L. R. Wilson, D. J. Mowbray, M. S.
Skolnick, M. Morifuji, M. J. Steer, I. A. Larkin, and M.
Hopkinson, {\it Phys. Rev. B} {\bf 57}, 2073 (1998); M. Hayne, J.
Maes, V. V. Moshchalkov, Y. M. Manz, O. G. Schmidt, and K. Eberl,
{\it Appl. Phys. Lett.} {\bf 79}, 45 (2001).

\bibitem{bufon04} C. C. B. Bufon {\it et al.}, manuscript in
preparation.

\bibitem{fock28} V. Fock, {\it Z. Phys. } {\bf 47}, 446 (1928).


\end{thebibliography}
\end{document}